\documentclass[a4paper,11pt]{article}
\usepackage[left=0.8in,right=0.8in,top=0.8in,bottom=0.8in]{geometry}
\usepackage{graphicx}
\usepackage{float}
\usepackage{amsmath,amssymb}
\usepackage[english]{babel}
\usepackage{physics}
\usepackage{hyperref}
\usepackage{authblk}
\usepackage[autostyle]{csquotes}
\newcommand{\be}{\begin{equation}}
\newcommand{\ee}{\end{equation}}
\newcommand{\bea}{\begin{eqnarray}}
\newcommand{\eea}{\end{eqnarray}}
\newcommand{\nn}{\nonumber}

\def\p{\partial}
\title{{\bf  Quantum chaos measures for Floquet dynamics}}
\author{{Amin A. Nizami}
\thanks{email: amin.nizami@ashoka.edu.in}}
\affil{\small Department of Physics, Ashoka University, Rajiv Gandhi Education City, Rai, NCR 131029, India\\
Orcid ID: 0000-0003-3902-5320}
\date{} 
\begin{document}
\maketitle

{\abstract  Periodically kicked Floquet systems such as the kicked rotor are a paradigmatic and illustrative simple model of chaos. For non-integrable quantum dynamics there are several diagnostic measures of the presence of (or the transition to) chaotic behaviour including the Loschmidt echo, autocorrelation function and OTOC. We analytically compute these measures in terms of the eigensystem of the unitary Floquet operator of driven quantum systems.  We use these expressions to determine the time variation of the measures for the quantum kicked rotor on the torus, for the integrable as well as the chaotic case.  For a simpler integrable variant of the kicked rotor, we also give a representation theoretic derivation of its dynamics.}

\section{Introduction} \label{intro}

The dynamics of periodically driven quantum systems is of physical interest for a number of reasons. These Floquet systems provide an arena for exploring chaotic and non-equilibrium quantum dynamics and give access to interesting new phases of matter away from thermal equilibrium such as time crystals, while also providing the opportunity to engineer and precisely control quantum states.

Such systems also provide the simplest avenues to study chaotic behaviour. Classical chaos is diagnosed through two characteristic features: {\it local instability} - an exponential divergence of nearby trajectories in phase space and {\it global mixing} - intuitively a strong delocalisation or spreading of concentrated initial conditions throughout phase space, a property which implies ergodicity. The Poincare-Bendixson theorem rules out chaotic behaviour in the phase plane, that is, for autonomous systems with one degree of freedom (d.o.f) - or equivalently a 2 dimensional phase space. Seeking chaos, one may next turn to the non-linear dynamics of two coupled  d.o.f (4 dimensional phase space, for example that of 2 coupled non-linear oscillators) or a {\it driven} system - this has a 3 dimensional phase space (1.5 d.o.f). Alternatively one may study discrete dynamical systems (classical or quantum maps) where quite simple models can show extremely rich dynamics. Delta function kicked driven systems, the subject of our study, are examples of this kind.

To find chaotic behaviour one has to look at many-body systems (or at least with more than one d.o.f) which are strongly interacting. As a result, perturbative methods are usually of not much help since the KAM theorem implies that integrable dynamics is robust to small perturbations (weak nonlinearities). It is only when appropriate coupling constants in the hamiltonian are tuned to suitably large values, thereby exiting the perturbative regime, that the KAM tori start to break apart and the phase space typically starts exhibiting expanding islands of chaotic behaviour as inferred by studying, for example, the Poincare sections. A quantum version of the KAM theorem has proved elusive, although see \cite{KAM}

For these reasons, most studies of chaotic behaviour are numerical.  In the quantum case, one quantizes a dynamical system which in its classical limit can show both integrable and chaotic behaviour, and looks at suitable diagnostic measures with qualitatively different features in the two cases \cite{chaos}. We may expect that for simpler systems with only a few d.o.f at least some of these measures of chaos may be calculated exactly, thereby giving an analytical handle to study the onset of quantum chaos by tuning parameters in the hamiltonian and looking for characteristic behaviour.

In this paper we revisit the quantum kicked rotor (QKR), a much studied simple model of classical and quantum chaotic behaviour \cite{CCIF, qkr}.  Also discussed will be the related models of the QKR on a torus as well as a linear (in momentum) integrable variant. The dynamics of the canonical QKR is encapsulated by the time dependent Hamiltonian:

\be
H(t)= H_0+H_{int}(t)=\frac{p^2}{2M}+ V(\theta) \sum_{p=1}^{\infty} \delta(t-pT)  
\ee
Here  $V(\theta)$ is periodic in $\theta$.  The choice $ V(\theta)=K\cos\theta$ generates the so-called standard or Chirikov map. This represents a particle moving on a circle, kicked periodically in time with a Dirac delta function potential, and with a non-linear spatial part which is also periodic and has a strength parameter ($K$).  Such a set-up can be realised experimentally \cite{exp} - for example, by suitably confining cold atoms/molecules kicked periodically by a pulsed laser. The (non-relativistic) motion of a particle in a circular accelerator boosted repeatedly through periodic energy kicks can be modelled similarly.

Classically, the integrated Hamilton's equations of motion of this system (with $ V(\theta)=K\cos\theta$ ) give the well-known Chirikov map: 
\be
\theta_{j}=\theta_{j-1}+(T/M) P_{j-1} \,\,\,\,,\,\,\,\,\,P_{j}=P_{j-1}+K \sin\theta_j \label{Chmap}\\
\ee
where $\theta_j,\,\, P_j$ denote the value of the observable right after the $j$th kick at $t=jT$. As the kicking strength, determined by the parameter $K$, is increased the system transitions from regular to chaotic behaviour. See \cite{revCh} for an overview of the QKR. 

As is well known, in the quantum case this model exhibits dynamical localisation - a one-dimensional variant of Anderson localisation - where the classical energy diffusion, as measured by $\ev{P^2}$, is suppressed after a certain time due to quantum interference effects \cite{CCIF}. This is seen to be a characteristic feature of quantized systems whose classical limit is chaotic - a quantum mechanical \enquote{softening} of classical chaotic effects. 

There are multifarious quantitative measures used for detecting chaotic behaviour when studying the quantized dynamics of classically non-integrable systems. As is well known, the state overlap given by the inner product $\braket{\psi(0)}{\psi(t)}$ can not be used since the linear unitary evolution of quantum states is norm-preserving and therefore shows no sensitive dependence on initial conditions\footnote{Other measures of state distinguishability in Hilbert space could still be used -  for example the relative state complexity \cite{YC} .}. An alternative is to detect sensitivity of the hamiltonian to small perturbations \cite{Peres}. This leads to the fidelity/Loschmidt echo studied in section \ref{chaos}. Also of much utility is the thermal out of time order correlator (OTOC) or chaos correlator first considered by \cite{Larkin} and more recently studied, from the perspective of holography and quantum information scrambling in black holes, as a chaos diagnostic \cite{ MSS}.

We will derive analytical expressions for several quantum chaos measures in terms of the eigensystem of the Floquet operator. These are known to show characteristically different behaviour in the regular and chaotic regime - for example, the Loschmidt echo and the thermal OTOC  typically have an exponential form up until the Ehrenfest time in the presence of chaos. One reason the chaos measures are useful is that one can study, numerically or analytically, the transition from regular to chaotic behaviour in the quantum domain. 

In section \ref{basics} , after a brief review of the salient features of the classical and quantum kicked rotor, we discuss how Floquet theory simplifies its study and formulate the eigen-equation for the Floquet operator in various forms. 
In section \ref{chaos} we analytically compute the autocorrelation function, OTOC, Loschmidt echo and Wigner function in terms of the eigensystem of the unitary Floquet operator of this system. We also determine the time variation of some of these measures for the toral QKR, for both the regular and chaotic cases. In section \ref{linear} we look at an integrable simpler variant of the kicked rotor first studied in \cite{GFP} and use group representation theory to solve for its dynamics. We conclude with a brief discussion in section \ref{discuss}.

\section{The kicked rotor: basics and Floquet theory} \label{basics}

The particle on a ring is among the simplest quantum systems and yet quite rich in its physics. The Hamiltonian for a free particle (mass $M$) on a ring (radius $R$) is
\be
H_0= \frac{p^2}{2M}=\frac{-\hbar^2}{2M R^2}\frac{\p^2}{\p \theta^2}
\ee
The energies and normalised eigenfunctions are:
\be
E_n=\frac{n^2\hbar^2}{2MR^2}\,,\,\,\,\, \psi_n(\theta)=\frac{e^{i n \theta}}{\sqrt{2 \pi}} \label{FEF}
\ee
and a general state can be expanded in a basis of these momentum/energy eigenstates ($P \ket{n}= \frac{n\hbar}{R} \ket{n}$)
\be
\Psi(\theta,t)=\sum_{n} \alpha_n e^{i(n\theta - \omega_n t)}
\ee

The Hamiltonian for a kicked rotor is
\be
H(t)= H_0+H_{int}(t)=\frac{p^2}{2M}+V(\theta) \sum_{p=1}^{\infty} \delta(t-pT)=\frac{-\hbar^2}{2M R^2}\frac{\p^2}{\p \theta^2}+V(\theta)\sum_{p=1}^{\infty} \delta(t-pT)
\ee
We will set $M=R=1$ and often use $V(\theta)=K\cos\theta$ which generates the quantum standard map.

The analysis of periodically driven quantum systems is greatly simplified by using Floquet theory. Given a hamiltonian
\be
H(t)=H_0 +\bar{V}(t)
\ee
where the interaction part is periodic in time- $\bar{V}(t+T)=\bar{V}(t)$, one decomposes the state at time $t$ in terms of the Floquet eigenstates $\ket{\psi_n(t)}$
\be
\ket{\psi(t)}=\sum_n c_n \ket{\psi_n(t)}
\ee
One can show using Floquet theory that 
\be
\ket{\psi_n(t)}=\exp(-\frac{i \epsilon_n t}{\hbar})\ket{\phi_n(t)}
\ee
where $\epsilon_n$ - defined modulo $2 \pi/\hbar$ are the quasi-energies and $\phi_n(t)$ - the Floquet modes - are periodic and need only be defined over one time period - $[0,T]$.\footnote{This is the analogue of the well known Bloch theorem for spatially periodic potentials.}

The Floquet operator is the single step time evolution operator (generating time evolution over a time period $T$) and will be denoted by $U_F(T)$. To solve for the state at time $t$:
\begin{itemize}
\item Diagonalise $U_F(T)$ to obtain eigenvalues: $\exp(-\frac{i \epsilon_n T}{\hbar})$, and eigenvectors: $\ket{\phi_n (0)}$.
\item The initial state is decomposed in the Floquet eigenbasis : 
\be
\ket{\psi(0)}=\sum_n c_n\ket{\phi_n(0)}
\ee

\item it is easy to show that the state at any time t is then given by
\be
\ket{\psi(t)}=\sum_n c_n \exp(-\frac{i \epsilon_n t}{\hbar})\ket{\phi_n(t)}
\ee

where
\be
c_n =\braket{\phi_n(0)}{\psi(0)}\,\,\, \ket{\phi_n(t)}= \exp(\frac{i \epsilon_n t}{\hbar})U(t) \ket{\phi_n(0)}
\ee
\end{itemize}
where in the above equation $t$ is measured $mod\, T$ using Floquet periodicity.

The dynamics of a kicked system comprises of free propagation alternating with localised kicks. One can thus look at the system stroboscopically, right after a kick, at a time $t=jT$.
Time evolution is generated by iterations of the Floquet Operator $U(t=jT)=U_F(T)^j$

\be
U_F(T;K)=U_V U_0(T), \,\,\, U_V= \exp(\frac{-i}{\hbar}V(\theta)), \,\,\, U_0(T)=\exp(\frac{-i}{2\hbar}P^2 T)
\ee
where periodic kicks alternate with free propagation. 

For the kicked rotor with $V(\theta)=K \cos \theta$ the matrix form of the Floquet operator, in the $H_0$ eigenbasis, is known to be \footnote {This follows easily from the generating function of the Bessel function

\be
\exp(i k \sin \theta)=\sum_{n=-\infty}^{\infty} J_n(k)\exp(i n \theta) \nn
\ee
}

\be
U_{nm}=\mel{n}{U(T;K)}{m}= \exp(\frac{i m^2 \hbar T}{2}) (-i)^{m-n}J_{m-n}(K/\hbar) \label{Umn}
\ee

Before proceeding, a few side remarks about the interesting form of this matrix. 

\begin{itemize}

\item Due to the Bessel function, away from the diagonal the magnitude of the elements decreases quite fast - it is effectively a banded matrix.
\item Except for a diagonal part (the exponential factor), this is a {\it Toeplitz} matrix \footnote {I thank Rajendra Bhatia for pointing this out.} - the matrix elements $A_{mn}$ are of the form $A_{m-n}$, that is, all elements on any fixed super/sub - diagonal are the same.  
\item  There are two natural time scales in this problem - the time period for motion around the circle and that of periodic forcing.  When the ratio of the associated angular frequencies is a rational number 
\be
\frac{\omega_0}{\omega}= \frac{\hbar T}{4 \pi M R^2}= \frac{p}{q}
\ee
we get the quantum resonance condition and it was shown by \cite{CSRes} that in this case the problem for the Floquet operator can be effectively reduced to one for a finite dimensional $q \times q$ matrix.
\item Consider the Euclidean group in 2 dimensions - $E(2)$,  generated by rotations $R_{\theta}$ (parametrised by angle $\theta$) about a fixed point and translations $T_{\vec{a}}$ (parametrised by $a$ and $\phi$ - giving the magnitude and direction of the translation vector $\vec{a}$) on a plane. A general element of this group can be written as $g=T_{\vec{a}} R_{\theta}$.  Its infinite dimensional unitary irreducible representations are known and given by
 \be
 \mathcal{U}_{nm}= e^{-i m \theta} e^{i (m-n) \phi}J_{m-n}(\lambda a) \label{Urep}
 \ee
 $\lambda$ is a continuous positive parameter labelling the representation and $m$ labels the basis states in a particular representation space. This is quite close structurally to our $U_{nm}$, again except for the diagonal part - we will use the above form to solve for a simpler variant of the kicked rotor in section \ref{linear}.
 
\end{itemize}

For the Floquet operator $U_F(T)$, we denote its eigenvalues by $\mu_n$,  corresponding eigenvectors by $\ket{\mu_n}$ and the eigenfunctions by $\mu_n (\theta)$.

The general time-evolved state (after $j$ kicks) is given by

\be
\ket{\psi(t)}=U_F^j \ket{\psi(0)}  =\sum_n c_n \mu_n^{j}\ket{\mu_n} 
\ee
with $c_n=\braket{\mu_n}{\psi(0)}$. The wave-function at time $t=jT$ is then given by:

\be
\Psi(\theta,t) =\sum_n c_n (\mu_n)^{j} \mu_n (\theta) \label{wfnp}
\ee
where
\be
\mu_n(\theta)= \sum_m e^{im\theta}\tilde{\mu}_n^{(m)}
\ee
and $\tilde{\mu}_n^{(m)}=\braket{m}{\mu_n}$ is the $m$th Fourier coefficient of the Floquet eigenfunction $\mu_n (\theta)$.

The momentum space wave function after $j$ kicks is (with $p=m \hbar$)
\be
\Psi(m,t=jT) = \sum_n \mu_n^j \braket{m}{\mu_n} \braket{\mu_n}{\psi(0)}= \sum_n c_n \, \mu_n^j \, \tilde{\mu}_n^{(m)}                                         \label{wfnm}
\ee
We will use these expressions in section \ref{chaos}.

\section{Measures of quantum chaos for driven systems} \label{chaos}
In this section we will derive expressions for a number of chaos diagnostic measures - namely the Loschmidt echo, autocorrelation function and OTOC - in Floquet systems in terms of the eigensystem of the associated Floquet Operator. Appendix C provides a related expression for the Wigner function. These expressions hold for general Floquet systems which are probed stroboscopically and not just delta-function kicked systems like the quantum standard map. We will finally use the expressions to obtain plots of the time variation of some of these measures for a finite dimensional variant of the QKR, defined on a toroidal phase space.

\subsection*{Loschmidt echo}

One way of diagnosing chaos in the quantum regime is to study the sensitivity of the hamiltonian (rather than the initial state) to perturbations, which may be due to small changes in its parameters. This was done first  by A. Peres \cite{Peres} and leads naturally to studying the time variation of the quantum fidelity or Loschmidt Echo \cite{LErev}. This quantity is important also in the study of quantum decoherence \cite{LED}.

Consider a fixed initial state $\ket{\psi(0)}$ at $t=0$ and evolution through two \enquote{neighbouring} hamiltonians $H$ and $H'=H+\delta H$ to respective states $\ket{\psi(t)}$ and $\ket{\psi'(t)}$. In particular the perturbation may be due to a variation of a parameter $K$ in the hamiltonian which controls the interaction strength (with $K'=K+\delta K$). The {\it fidelity} of the evolution is defined by

\be
L(t)=|\braket{\psi_{K'}(t)}{\psi_K(t)}|^2
\ee
and naturally measures the deviation, after a time $t$, induced by the small initial perturbation. This can also be interpreted as forward evolution for time $t$ of the initial state $\ket{\psi(0)}$ by the hamiltonian $H_K$  followed by backward evolution for time $t$ by $H_{K'}$ (imperfect time reversal) and then taking the overlap with the initial state. Thus viewed, the same quantity is called the Loschmidt echo. This function typically has an exponential/power-law decay in the chaotic/regular regime \cite{Peres,LED}. For the QKR, it has been studied in \cite{AL} . 

To calculate the Loschmidt echo/fidelity for a periodically driven system we will take the initial state to be $\ket{\psi(0)}$  - which could be the free eigenstate $\ket{m}$ - and then consider forward/backward evolution through $j$ kicks. We have the echo operator $E_j=U_F^{\dagger}(T,K')^{j}U_F(T,K)^j$ and the Loschmidt echo is defined in terms of the expectation value of the echo operator by $L_j=|\ev{E_j}|^2$.


For evolution in a general Floquet system for time $t$ which includes $j$ full time periods,  
\be
U(t)=U_0(t-jT)U_K U_0(T)........U_K U_0(T)U_K U_0(T)
\ee
with $j$ factors of $U_K$ and,
\be
\ev{E}_{j}=\mel{\psi(0)}{U_F^{j\dagger}(T,K')U_F^j(T,K)}{\psi(0)}
\ee

Using Floquet theory, one can get an alternative expression in terms of the eigen-data of the Floquet operator .Taking  the initial state $\ket{\psi(0)}=\ket{n}$ and after introducing the completeness relation $1=\sum_m \ket{\mu_m}\bra{\mu_m}$ in the above expression, it follows that

\be
\ev{E_j}=\sum_{m} \tilde{\mu}_{m}^{(n)}(K') \tilde{\mu}_m^{(n)*}(K) \big(\mu_m(K)\mu_{m}^*(K')\big)^{j} \nn
\ee
 Here $\mu_n(K)$ and $\ket{\mu_n(K)}$ are the eigenvalues and eigenvectors of the Floquet operator $U_F(T;K)$.  This gives the required expression for the Loschmidt Echo in terms of the Floquet eigen-data, $L_j=|\ev{E_j}|^2$.

\subsection*{Autocorrelation function}

To study operator growth and complexity in a chaotic system, one natural object to study is the autocorrelation function $A(t)=Tr \big(P(t)P(0)\big)$ where $P(t)$ is the Heisenberg picture momentum operator at time $t$. Probing the Floquet dynamics stroboscopically at $t=jT$, we have  $P_j= U_F^{\dagger j}P_0 U_F^j $. Calculating the trace in the Floquet eigenbasis gives
\be
A_j= \sum_n \mel{\mu_n}{U_F^{\dagger j}P_0 U_F^j P_0}{\mu_n}
\ee
Inserting the completeness relation $1=\sum_k \ket{\mu_k}\bra{\mu_k}$ in between each operator yields, after simplifying, the sought relation
\be
A_j= \sum_{m,n} |\mel{\mu_m}{P_0}{\mu_n}|^2 (\mu_n^* \mu_m)^j
\ee


\subsection*{OTOC}

Along the same lines as above, one may compute the OTOC 
\be 
C(t)=-\mel{\psi(0)}{[P(t),P(0)]^2}{\psi(0)}
\ee
starting with 
\be
[P_j,P_0]\ket{\psi(0)}= \sum_{m,n} (\mu_n^* \mu_m)^j \mel{\mu_n}{P_0}{\mu_m} \big( \mel{\mu_m}{P_0}{\psi(0)} -P_0 \braket{\mu_m}{\psi(0)}\big)\ket{\mu_n}
\ee
The norm of the above state gives the expression for the OTOC, with the initial state $\ket{\psi(0)}=\ket{n}$
 \be \label{Cj}
 C_j= \hbar^2 \sum_m (n-m)^2 |\mel{m}{P_j}{n}|^2 
 \ee
with
\be
\mel{m}{P_j}{n}=\sum_{k,l} (\mu_k \mu_{l}^*)^j \tilde{\mu}_{k}^{(m)} \tilde{\mu}_{l}^{(n)*} \mel{\mu_k}{P_0}{\mu_l} 
\ee 
The QKR OTOC was studied numerically in \cite{notoc}. 

\subsection{Illustrative example: The toral QKR} \label{plots}
 We will now compute these chaos measures in a particular model - the QKR on a torus, to illustrate the utility of the above expressions. This model is obtained by further imposing periodic boundary condition on $p$. The Floquet operator in the $q$ basis takes the form

  \begin{align}
    (U_F)_{nn'} = \frac{1}{N}e^{\frac{iN\kappa}{2\pi}\cos(2\pi\frac{(n+\alpha)}{N})}\sum_{m=0}^{N-1}e^{\frac{-\pi i (m+\beta)^2}{N}} e^{\frac{2\pi i}{N}(m+\beta)(n-n')}
\end{align}

Here $N$ is the dimension of the Hilbert space, $\kappa$ is the coupling constant and the indices $n, n'$ run from $0$ to $N-1$. The breaking of parity and time-reversal symmetries is determined by non-zero $\alpha$ and $\beta$ parameters, respectively. For any finite $N$, the eigenvalues and eigenvectors of this matrix can be determined and the above expressions used to compute the measures. 

The time variation of the autocorrelation function, OTOC and Loschmidt echo is plotted in the figures below. We choose two sets of values for the couplings, encompassing the regular and chaotic cases. OTOCs for various quantum maps have been computed in \cite{G1}. We notice from the figures that in the chaotic case both the autocorrelation function and Loschmidt echo decay quickly whereas the OTOC increases and reaches its saturation value. 
Compared to this, the saturation values are noticeably different in the regular (weak coupling case) for both the autocorrelation function and the OTOC whereas the Loschmidt echo in this case shows characteristic oscillatory behaviour.
The sharp decay of the autocorrelation function in the chaotic case is expected on physical grounds. $A_j$ can be interpreted as the Frobenius inner product $(P_0|P_j)$ in the space of operators. When the dynamics is stochastic $P_j$ very quickly becomes uncorrelated with $P_0$ even for a small number of kicks $j$ thus explaining the observed behaviour.

\begin{figure}[H]
    \centering
    \includegraphics[height = 5.5 cm, width = 7.2cm]{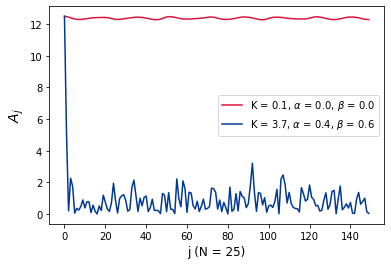}
    \includegraphics[height = 5.5 cm, width = 7.2cm]{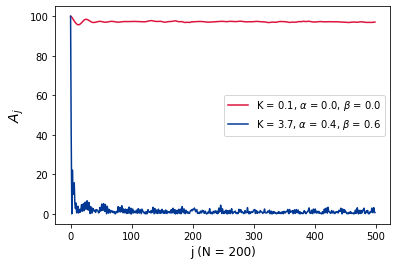}
    \caption{Autocorrelation function: The plots of $A_j=Tr(P_j P_0)$ vs time in the regular and chaotic regime are displayed for two different system sizes: $N=25$ (left) and $N=200$ (right).}
    \label{fig:1}
\end{figure}

\begin{figure}[H]
    \centering
    \includegraphics[height = 5.5 cm, width = 7.2cm]{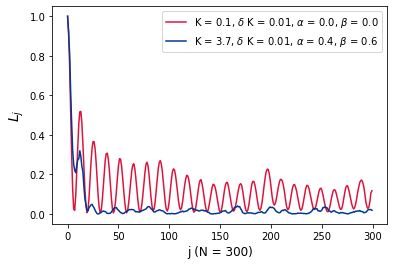}
    \includegraphics[height = 5.5 cm, width = 7.2cm]{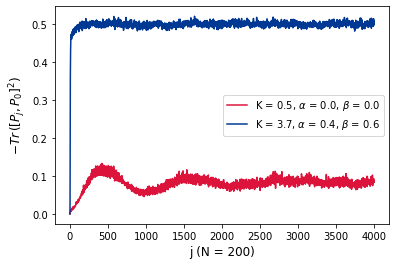}
    \caption{Loschmidt Echo vs. time (left): Note the characteristic oscillatory behaviour in the regular regime. OTOC (right): The figure shows the growth and saturation of the OTOC $C_j=-Tr[P_j,P_0]^ 2$ in the regular and chaotic cases.}
    \label{fig:2}
\end{figure}

\section{A simpler kicked rotor} \label{linear}
In this section we look at a variant of the kicked rotor which is linear rather than quadratic in the momentum, the hamiltonian being:

\be
H(t)=\alpha p+K\cos\theta \sum_{p=1}^{\infty} \delta(t-pT)
\ee

This system was studied in \cite{GFP}. It is an integrable system and its quasi energies and eigenfunctions have been analytically determined, see \cite{GFP} for details and also how the classical and quantum dynamics is distinguished for rational vs. irrational values of $\alpha$. Here we give a representation theoretic derivation of its quantum dynamics by solving for the unitary evolution operator for $j$ kicks.

The free part of the hamiltonian $H_0=\alpha P$ has eigenvalues $E_n= n \alpha \hbar$ whereas the eigenfunctions (which we use as a basis) are the same as in eq. \eqref{FEF} above. The Floquet operator corresponding to the hamiltonian above is:
\be
U_F(T)=\exp(-\frac{i K \cos \theta}{\hbar}) \exp(-\frac{i \alpha T P}{\hbar})
\ee
which, in the momentum eigenbasis, has the matrix form:
\be
(U_F)_{nm}= \exp(-\frac{i \alpha T m}{\hbar}) i^{m-n}J_{m-n}(K/\hbar)
\ee

Compare this with the form of the  irreducible infinite dimensional unitary matrix representations of the Euclidean group in 2 dimensions - $E(2)$ - given in eq. \eqref{Urep}. With $\theta= \alpha T/\hbar$, $\phi=\pi/2$, $a=K/\hbar$, and $\lambda=1$, we can identify the two expressions. The realisation that the Floquet operator generating quantum dynamics forms a representation of the symmetry group $E(2)$ is helpful in solving this problem. This is because the geometric group action of $E(2)$ in the plane:
 
 \be
 (R_2, \vec{a}_2)(R_1,\vec{a}_1)=(R_2 R_1, R_2 \vec{a}_1+\vec{a}_2)
 \ee
can be leveraged together with the group representation property that enables to write products of several $U$'s as a single $U(R, \vec{A})$. We get for the time evolution operator after $j$ kicks
\be
U_j=\big(U_F(R_{\theta},\vec{a})\big)^j = U_F\big(R_{\theta}^j, \vec{A}_j\big)
\ee
where
\be
\vec{A}_j= (R_{\theta}^{j} + R_{\theta}^{j-1}+R_{\theta}^{j-2}+...+R_{\theta}   )\vec{a}
\ee
Here $R_{\theta}$ is the orthogonal rotation matrix (counterclockwise rotation by angle $\theta$) and $\vec{a}=(0, \, a)^T$. The vector $\vec{A}_j$ has magnitude $A_j = a\sin(j \theta/2)/\sin(\theta/2)$ and angle $\Phi_j=\big(\pi+(j+1)\theta\big)/2$ with the $x$-axis. This geometrically corresponds to an alternating sequence of $j$ rotations (each of angle $\theta$) and translations (parallel to the $y$ axis, each of magnitude $a$) in the plane. Thus we can write the full final form of the time evolution operator after $j$ kicks as:

\be \label{Uj}
(U_j)_{nm}= \exp(-\frac{i }{2\hbar}(j+1)m\,\alpha T) i^{m-n} J_{m-n}\Big(\frac{K\sin(j\alpha T/2\hbar )}{\hbar \sin(\alpha T/2\hbar )}\Big)
\ee

This completely determines the dynamics. Note how the use of the group representation condition gives the above compact form. Measures like the Loschmidt echo can be now computed. For example the echo operator can be evaluated using a similar method: after $j$ translations and rotations, we subsequently perform $j$ inverse transformations which partially undo the first sequence. The Loschmidt echo for $j$ kicks, with the initial state being any $H_0$ eigenstate, can be calculated to be given by the simple expression
\be
L_j= J_0^2\Big( \frac{|\delta K| \sin(j \theta_0)}{\hbar \sin \theta_0}\Big), \,\,\,\,\, \theta_0 = \alpha T/2\hbar
\ee
Of course, in this case no chaotic behaviour is expected - this function oscillates periodically. 
The autocorrelation function is similarly computed to be 
\be
A_j= Tr (P_j P_0) = \hbar^ 2 \sum_{m,n} \,\, mn\, J^ 2 _{m-n} \Big(\frac{K \sin (j\theta_0)}{\hbar \sin (\theta_0)} \Big).
\ee

The OTOC is given by eq.  \eqref{Cj} where the matrix element can be evaluated using the closed form expression in eq. \eqref{Uj} to be (an overall phase factor has been neglected below):
\be
\mel{m}{P_j}{n}=\hbar^ 2 \sum_l l\, J_{m-l}\Big(\frac{K \sin (j\theta_0)}{\hbar \sin (\theta_0)}\Big) J_{n-l}\Big(\frac{K \sin (j\theta_0)}{\hbar \sin (\theta_0)}\Big). 
\ee
In the limit of a high frequency drive ($T\rightarrow0$), the early-time behaviour ($j$ dependence) of the OTOC can be extracted: $C_j \sim K^2 j^2$.

 The Wigner function for this system (see appendix \ref{WF}), in the initial state $\ket{n}$, can also be computed - starting from eq. \eqref{W} - in the form of a Fourier mode expansion
\be
W_n(\theta, p_l;t)=\frac{(-1)^{n-p_l}}{\pi} \sum_r e^{2ir\theta} J_{n-p_l-r}\Big(\frac{K\sin(j \theta_0)}{\hbar \sin \theta_0 }\Big) J_{n-p_l+r}\Big(\frac{K\sin(j \theta_0)}{\hbar \sin \theta_0}\Big).
\ee

Note that the method used in this section can be applied whenever the unitary Floquet operator of the system forms a group representation, with the group having a natural geometric action on a space. 

\section{Discussion} \label{discuss}

We have studied various aspects of Floquet quantum dynamics for kicked systems. Our main results include explicit expressions for various quantum chaos measures in terms of the Floquet operator eigensystem and a representation-theoretic derivation of the dynamics of an integrable kicked rotor.  For the toral QKR, we plotted and noted the characteristically different behaviour of the different chaos measures in the regular vs chaotic domain. Although we did not solve the Floquet eigen-problem analytically for the QKR, the results obtained should be helpful also in numerical studies of the various diagnostic measures of quantum chaos in other driven systems. It is possible that a deeper understanding of QKR dynamics can be obtained also from the alternative forms of the eigen-equation in appendix A.  The representation theoretic treatment for the integrable variant of the kicked rotor (section \ref{linear}) was based on the insight that its Floquet operator forms a representation of the Euclidean group of planar motion $E(2)$. For the QKR (standard map), as noted in section \ref{basics}, the form of the Floquet matrix is very similar - see eq.\eqref{Umn}. It would be quite interesting if representation theoretic methods can be wielded also in this more complex chaotic case to gain further insights into its dynamics.

One measure of chaos that we did not study here is quantum circuit complexity \cite{qch}. In the chaotic regime this is expected to increase linearly with time before saturating  and it should be insightful to calculate it for the simple models considered in this paper. The study of Krylov Complexity for Floquet dynamics was recently undertaken in \cite{KC}. It should also be interesting to study other simple Floquet systems where similar analytical computations are feasible, and the chaos measures can be computed and used to investigate the classical to quantum as well as the regular to chaotic transition.

\section*{Acknowledgements} I thank Arpan Bhattacharya for discussions and suggesting some useful references. I would also like to thank Rajendra Bhatia for useful discussions and Ankit Shrestha for help with the plots in section \ref{plots}.

\section*{Appendix A: Forms of the Floquet operator eigen-equation} \label{EE}

The  eigen-equation for the Floquet operator is:
\be
U\ket{\mu_m}=\mu_m \ket{\mu_m}
\ee
In the $H_0$ eigenbasis, this reads \big(with $\tilde{\mu}^{(n)}_m= \braket{n}{\mu_m}$ \big)
\be
\sum_{m'} \exp(\frac{i m'^2 T}{2\hbar}) (-i)^{m-m'}J_{m-m'}(K/\hbar) \tilde{\mu}^{(n)}_{m'}=\mu_m \tilde{\mu}^{(n)}_m  \label{UE}
\ee
an eigen-equation for an infinite dimensional unitary matrix.  This eigen-equation can be transformed to one for a Toeplitz hermitian matrix, as discussed in the next appendix. 

The Floquet operator is usually presented as a matrix in the momentum eigenbasis as above. It is possible, however, to write it in different equivalent forms. For example, in the position basis the eigen-equation reads:
\be
\Bigg(\exp(\frac{-i}{\hbar} K \cos \theta) \exp(\frac{i\hbar T}{2}\p_\theta^2) - \mu \Bigg) \mu(\theta)=0 \label{ODE}
\ee
an infinite order linear ODE. Quantisation of $\mu$ would be imposed by requiring the wave function $\mu(\theta)$ to be single-valued. A Discrete Fourier transform of $\mu(\theta)$ leads to the matrix eigen-equation eq. \eqref{UE} above.  

An integral Fourier transform of the above equation to frequency space leads to (after using the convolution theorem for Fourier transforms) the relation

\be
\mu^{-1}  \exp(-i \hbar T \omega ^2 /2) \tilde{\phi}_\mu(\omega)=(\tilde{f}*\tilde{\phi}_\mu)(\omega)
\ee
where $*$ denotes integral convolution,  $f(\theta)=\exp(i K \cos \theta /\hbar)$ - whose Fourier tansform gives a Bessel function. We therefore get the following (type 2 Fredholm) linear integral equation:

\be
 \sqrt{2 \pi} \int d\nu \,\,i^{\omega-\nu} J_{\omega-\nu}(K/\hbar) \exp(i \hbar T \omega ^2 /2)\tilde{\phi}_\mu(\nu)= \mu^{-1}\tilde{\phi}_\mu(\omega)  \label{IE}
\ee
These equations should be useful in obtaining more information about the spectrum of the QKR Floquet operator.

\section*{Appendix B: Unitary to Hermitian Mapping of the Floquet Operator eigen-problem}
It was shown by \cite{UH} that the eigen-problem for the Floquet operator can be mapped to an equivalent problem of a particle hopping on a 1D lattice.  A unitary operator $U = \exp(i h)$ can be written in terms of a hermitian operator $H$ in the alternative way (Cayley transform):
\be
U=\frac{1+i H}{1-i H},\,\,\,\, H=\tan(h/2)
\ee
One can use this to map the eigen-problem for the unitary Floquet operator to one for an associated hermitian operator. Let $\exp(-i \lambda)$, $\phi_{\lambda}$ be the eigenvalue and corresponding eigenvector of $U_F(T)$. Define:
\be
u_{\lambda}=(\exp(i \lambda)U_f+1)  \phi_{\lambda}
\ee
The Floquet operator is $U_F(T)=U_V U_0(T)$, if we absorb the $U_0$ part in the eigenfunction and use the Cayley transform of $U_V$, the eigen-equation for $U_F(T)$ can be cast in the form
\be
\tan(\frac{T H_0}{2\hbar}-\frac{\lambda}{2})u_{\lambda}=\tan(\frac{V(\theta)}{2\hbar}) u_{\lambda} 
\ee

In terms of the Fourier modes of $u_{\lambda}$, one can write the eigenvalue equation in the $H_0$ eigenbasis
\be
\sum_n f_{m-n}u_{\lambda}^n=\tan(\frac{m^2 T}{4\hbar }-\frac{\lambda}{2})u_{\lambda}^m
\ee
where $f_k$ is the $k$-th Fourier coefficient of $f(\theta)=\tan(\frac{V(\theta)}{2\hbar})$:
\be
f_k=\frac{1}{2\pi}\int_0^{2\pi} d\theta \tan(\frac{V(\theta)}{2\hbar}) e^{-i k \theta}
\ee

Note that we have transformed the eigen-equation of a unitary matrix to one of a hermitian matrix \footnote{We find it convenient to write the above equation in this form - it is usually cast in a form describing the hopping amplitude of a particle in a 1D tight binding lattice model \cite{UH}}. For the family of periodic potentials $V(\theta)=-2 \hbar\, tan^{-1}(W(\theta))$, the Fourier coefficients are simple ($f_k=-W_k$). The eigenvalues, in such cases, may often be calculated analytically - for example \cite{UH} for $W(\theta)=k \cos \theta-E$, the so called Lloyd's model - it gives a nearest neighbour hopping tight binding model - we have $\lambda_m \propto m^2 \,\,(mod\, 2\pi)$.

Note also that the hermitian matrix is now Toeplitz. Although we do not pursue this further here, Toeplitz matrices have special properties  \cite{Toep} which are of use in the study of this problem. For example, to calculate sums over functions of eigenvalues (such as a spectral form factor) of an infinite dimensional hermitian Toeplitz matrix we may use Szego's theorem

\be
\lim_{n\rightarrow \infty} \frac{1}{n} \sum_{k=1}^{n}  F(\lambda_{k,n})=\frac{1}{2 \pi} \int_0^{2\pi} d\theta \, F(f(\theta)) 
\ee

Here $F$ is a function of the eigenvalues $\lambda_{k,n}$,  and $f(\theta)$ is the {\it symbol} of the Toeplitz matrix $f_{mn}$ - it is the (periodic) function whose Fourier coefficients are the elements of the Toeplitz matrix.

\section*{Appendix C: Wigner function} \label{WF}
The Wigner function \cite{wigner} is a quasi-probabilty distribution in phase space, and is often used in semiclassical analysis. It makes the connection to the classical limit particularly transparent. The \enquote{volume} of its negative part can be used to quantify how non-classical a state is \cite{KZ} and it is used as a chaos diagnostic as well \cite{wfn1}. See \cite{wfn3} for a numerical study of the QKR Wigner function. It also provides a route to an alternative phase space formulation of quantum mechanics \cite{GM}. For a review of its main properties and uses see \cite{wfn1, wfn2}. 
It is defined (in terms of position space wave functions) by:

\be
W(x,p;t)=\frac{1}{\pi \hbar}\int dy\, \Psi(x+y,t)\Psi^* (x-y,t) \exp(-2 i p y/\hbar)
\ee

For a Floquet system, using the explicit form of the wave function after $j$ kicks - eq.\eqref{wfnp}, the integral can be performed giving the following form in terms of the Floquet eigendata:
\be
W(\theta, p_l;t)=\frac{1}{\pi \hbar}\sum_{m,n,n'} c_n \mu_n^{j}\tilde{\mu}_n^{(2p_l-m)} \,e^{2i (m-p_l) \theta}\,  \tilde{\mu}_{n'}^{(m)*}\mu_{n'}^{*j} c_{n'}^* 
\ee
where $p_l$ is an integer labelling the momentum eigenvalue $p=\hbar p_l$.

Alternatively one may start with the Wigner function defined in terms of the momentum space wave function. Adapted to the case of a discrete momentum spectra \cite{wfn1} this definition is
\be
W(\theta, p_l;t)= \frac{1}{\pi} \sum_m  \Psi(p_l+m,t)\Psi^* (p_l-m,t) \exp(2 i m \theta) \label{W}
\ee
where the momentum space wave function $\Psi(n,t)$ is given by \eqref{wfnm}. This gives the same expression for the Wigner function as above.

\end{document}